\newif\ifjournal  % JOURNAL SUBMISSION?

\ifjournal
  \documentclass[Afour,sageh,times]{sagej}

  \usepackage[colorlinks,bookmarksopen,bookmarksnumbered,citecolor=red,urlcolor=red]{hyperref}

  \newcommand\BibTeX{{\rmfamily B\kern-.05em \textsc{i\kern-.025em b}\kern-.08em
T\kern-.1667em\lower.7ex\hbox{E}\kern-.125emX}}

\else
  \documentclass[twoside,leqno,twocolumn]{article}
  \usepackage{orcidlink}

  \usepackage{booktabs}
  \usepackage{caption}

  \usepackage{natbib}
  \setcitestyle{authoryear, square, yysep={;}}
  \setcitestyle{aysep={}}

\fi

\newcommand{\revision}[1]{#1}

\usepackage{algorithm}
\usepackage[noend]{algpseudocode}
\usepackage{cleveref}
\usepackage[cachedir=minted-cache,frozencache]{minted}
\usepackage{tikz}
  \usetikzlibrary{positioning}
\usepackage{xcolor}
\usepackage{xspace}

\newcommand{\code}[1]{\texttt{#1}}
\newcommand{\dbscan}{\textsc{Dbscan}\xspace}
\newcommand{\hdbscan}{\textsc{Hdbscan*}\xspace}
\newcommand{\eps}{\varepsilon}
\newcommand{\fdbscan}{\textsc{FDbscan}\xspace}
\newcommand{\fdbscandense}{\textsc{FDbscan-DenseBox}\xspace}
\newcommand{\find}{\textsc{Find}\xspace}
\newcommand{\minpts}{\textit{minPts}\xspace}
\newcommand{\unionfind}{\textsc{Union-Find}\xspace}
\newcommand{\union}{\textsc{Union}\xspace}
\newcommand\ttl[1]{\texttt{#1}}

\definecolor{RedOrange}{HTML}{FF4433}
\definecolor{Cerulean}{HTML}{007BA7}
\definecolor{Plum}{HTML}{DDA0DD}
\definecolor{OliveGreen}{HTML}{808000}
\newcommand{\fix}[1]{ {\color{Cerulean}#1}}

\begin{document}

\ifjournal
  \runninghead{Prokopenko et al}
\fi

\makeatletter
\def\blfootnote{\xdef\@thefnmark{}\@footnotetext}
\makeatother

% \title{A challenge is all you want: advances in ArborX to support exascale applications\fix{not happy with the title}}
\title{Advances in ArborX to support exascale applications}

\ifjournal
  \author{%
    Andrey Prokopenko\affilnum{1},
    Daniel Arndt\affilnum{1},
    Damien Lebrun-Grandi\'e\affilnum{1},
    Bruno Turcksin\affilnum{1},
    Nicholas~Frontiere\affilnum{2},
    J.D.~Emberson\affilnum{2},
    Michael~Buehlmann\affilnum{2}
  }
  \affiliation{%
    \affilnum{1}Oak Ridge National Laboratory, Oak Ridge, TN 37831, USA\\
    \affilnum{2}CPS Division, Argonne National Laboratory, Lemont, IL 60439, USA
  }
  \corrauth{Andrey Prokopenko,
  Oak Ridge National Laboratory,
  1 Bethel Valley Rd
  Oak Ridge, TN 37831
  USA.
  }
  \email{prokopenkoav@ornl.gov}
\else
  \author{
    Andrey Prokopenko\thanks{Oak Ridge National Laboratory}\enskip\orcidlink{0000-0003-3616-5504},
    Daniel Arndt\footnotemark[1]\enskip\orcidlink{0000-0001-8773-4901},
    Damien Lebrun-Grandi\'e\footnotemark[1]\enskip\orcidlink{0000-0003-1952-7219},
    Bruno Turcksin\footnotemark[1]\enskip\orcidlink{0000-0003-0103-888X},
    \\Nicholas Frontiere\thanks{CPS Division, Argonne National Laboratory}\enskip\orcidlink{0009-0005-8598-4292},
    J.D. Emberson\footnotemark[2]\enskip\orcidlink{0000-0003-1406-0744},
    Michael Buehlmann\footnotemark[2]\enskip\orcidlink{0000-0002-8469-4534}
  }
  \date{}
\fi

\ifjournal
\begin{abstract}
  ArborX is a performance portable geometric search library developed as part of
the Exascale Computing Project (ECP). In this paper, we explore a collaboration
between ArborX and a cosmological simulation code HACC. Large cosmological
simulations on exascale platforms encounter a bottleneck due to the in-situ
analysis requirements of halo finding, a problem of identifying dense clusters
of dark matter (halos). This problem is solved by using a density-based DBSCAN
clustering algorithm. With each MPI rank handling hundreds of millions of
particles, it is imperative for the DBSCAN implementation to be efficient. In
addition, the requirement to support exascale supercomputers from different
vendors necessitates performance portability of the algorithm. We describe how
this challenge problem guided ArborX development, and enhanced the performance
and the scope of the library. We explore the improvements in the basic
algorithms for the underlying search index to improve the performance, and
describe several implementations of DBSCAN in ArborX. Further, we report the history of
the changes in ArborX and their effect on the time to solve a representative
benchmark problem, as well as demonstrate the real world impact on production end-to-end
cosmology simulations.

\end{abstract}
\fi

\ifjournal
\keywords{geometric search, clustering, Kokkos, GPU, DBSCAN, cosmology}
\fi

\maketitle

\unless\ifjournal
  ArborX is a performance portable geometric search library developed as part of
the Exascale Computing Project (ECP). In this paper, we explore a collaboration
between ArborX and a cosmological simulation code HACC. Large cosmological
simulations on exascale platforms encounter a bottleneck due to the in-situ
analysis requirements of halo finding, a problem of identifying dense clusters
of dark matter (halos). This problem is solved by using a density-based DBSCAN
clustering algorithm. With each MPI rank handling hundreds of millions of
particles, it is imperative for the DBSCAN implementation to be efficient. In
addition, the requirement to support exascale supercomputers from different
vendors necessitates performance portability of the algorithm. We describe how
this challenge problem guided ArborX development, and enhanced the performance
and the scope of the library. We explore the improvements in the basic
algorithms for the underlying search index to improve the performance, and
describe several implementations of DBSCAN in ArborX. Further, we report the history of
the changes in ArborX and their effect on the time to solve a representative
benchmark problem, as well as demonstrate the real world impact on production end-to-end
cosmology simulations.

\fi

\blfootnote {%
% ORNL disclaimer
This manuscript has been authored by UT-Battelle, LLC, under contract
DE-AC05-00OR22725 with the U.S. Department of Energy. The United States
Government retains and the publisher, by accepting the article for
publication, acknowledges that the United States Government retains a
nonexclusive, paid-up, irrevocable, world-wide license to publish or reproduce
the published form of this manuscript, or allow others to do so, for United
States Government purposes.
}

\section{Introduction}\label{s:introduction}

ArborX is a performance portable geometric search library \citep{arborx2020}.
ArborX was developed as part of the Exascale Computing Project (ECP)
\citep{ecp}, a multi-year US Department of Energy (DOE) program. The ECP aimed
to provide an exascale computing ecosystem for DOE mission-critical
applications from many different domains, including cosmology, combustion,
material science, and additive manufacturing.

The complexity and the size of the DOE applications requires enormous computational
resources. DOE funded supercomputers, such as Frontier \citep{frontier},
Perlmutter \citep{perlmutter}, and Aurora \citep{aurora}, differ in their hardware
architectures, providing GPU accelerators from distinct vendors (AMD, Nvidia, and Intel, respectively).
Given that more than 95\% of the performance of each system is coming from these
accelerators, it is critical for a software library to be performance portable.

\iffalse
The ECP's research focus areas included Hardware and Integration (HI), Software
Technologies (ST), and Application Development (AD). In the ST area, ECP brought
together hardware vendors and library developers as one community to help to
exploit the full performance of the new hardware architectures. More than 70
unique software products were developed during the course of the project, with the
goal to support the upcoming supercomputers.
\fi

The primary goal of ArborX is to support DOE applications at scale.
Initially developed within the DataTransferKit library
\citep{datatransferkit}, ArborX became a standalone library in early 2019 when
it became clear that many other applications could benefit from its
functionality. Since then, ArborX experienced a significant growth in the
number of users and the diversity of use cases. ArborX now includes a wide
variety of algorithms that depend on the spatial proximity in the data: range
and nearest searches, clustering, ray tracing, and interpolation.

In this paper, we explore a collaboration of ArborX with a selected exascale
cosmology application that occurred over the 2020 -- 2023 period.

The Hardware/Hybrid Accelerated Cosmology Code (HACC; \cite{hacc,crkhacc}) is
an extreme-scale cosmological simulation framework.
The HACC challenge problem includes simulations with over ten trillion particles to
produce the most detailed synthetic sky maps ever made. In-situ analysis is
crucial, as these simulations would generate more than 100PB of data.
Consequently, the analysis suite has to scale, along with the code. ArborX
helps to address computational bottlenecks in several analysis components. In
this paper, we will focus on one such area: identification of clusters within
the full particle set.

The paper is organized as follows. We present the importance and the impact of
ArborX improvements on the performance of the analysis code in HACC in
\Cref{s:results}. We provide the background information on the algorithm and
ArborX library in \Cref{s:background}. We then describe performance and
algorithm improvements made in ArborX to achieve these results in
\Cref{s:updates}. Finally, we provide directions for further growth
in~\Cref{s:conclusions}.

\section{ArborX Impact on Production Simulations with HACC}\label{s:results}

\begin{figure}
    \centering
    \includegraphics[width=\columnwidth]{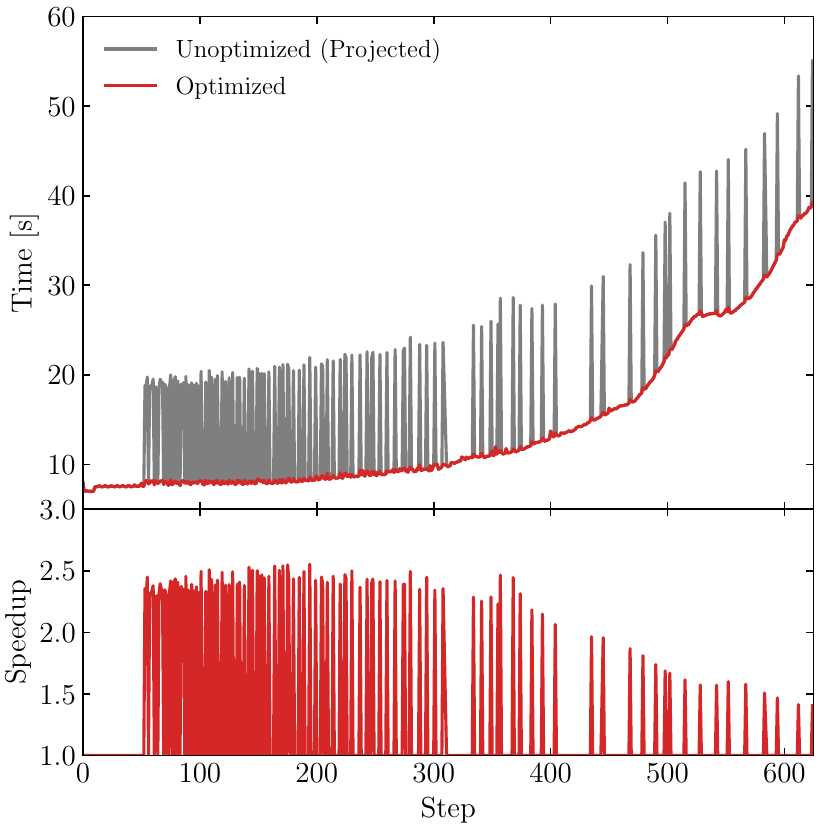}
    \caption{ Visualization of the performance impact of ArborX on analysis steps for a production gravity-only cosmology simulation.  }\label{f:hacc-speedup}
\end{figure}

Identification of halos (regions with a high density of dark matter particles)
is one of the most important analysis steps in cosmology. A common technique
for such identification is to use density-based clustering techniques FOF
(Friends-of-Friends), which is a specific case of a general \dbscan
(Density-based Spatial Clustering of Applications with Noise) algorithm
\citep{ester1996}. DBSCAN relies on the quick identification of particle
neighborhoods, grouping together points that are closely packed together while
marking the points in the low-density regions as noise (outliers).

Here, we discuss the capability impact of incorporating \dbscan algorithm
implementation developed in ArborX
within HACC by exploring a suite of full-scale end-to-end cosmology
simulations. The developments in ArborX required to achieve this impact are
described in \Cref{s:updates}.

The results were measured on the \emph{Summit} supercomputer using
256 nodes, each equipped with 6 Nvidia V100 GPUs. The domain volume is (576
Mpc/h)$^3$. We present results from both a gravity-only simulation, featuring
$N=2304^3$ particles, and a companion hydrodynamic simulation that evolved
twice as many particles to account for both dark matter and gas particles
individually. Both simulations trace the evolution from the early universe
(over 13 billion years ago) to the present.

\begin{figure}
    \centering
    \includegraphics[width=0.91\columnwidth]{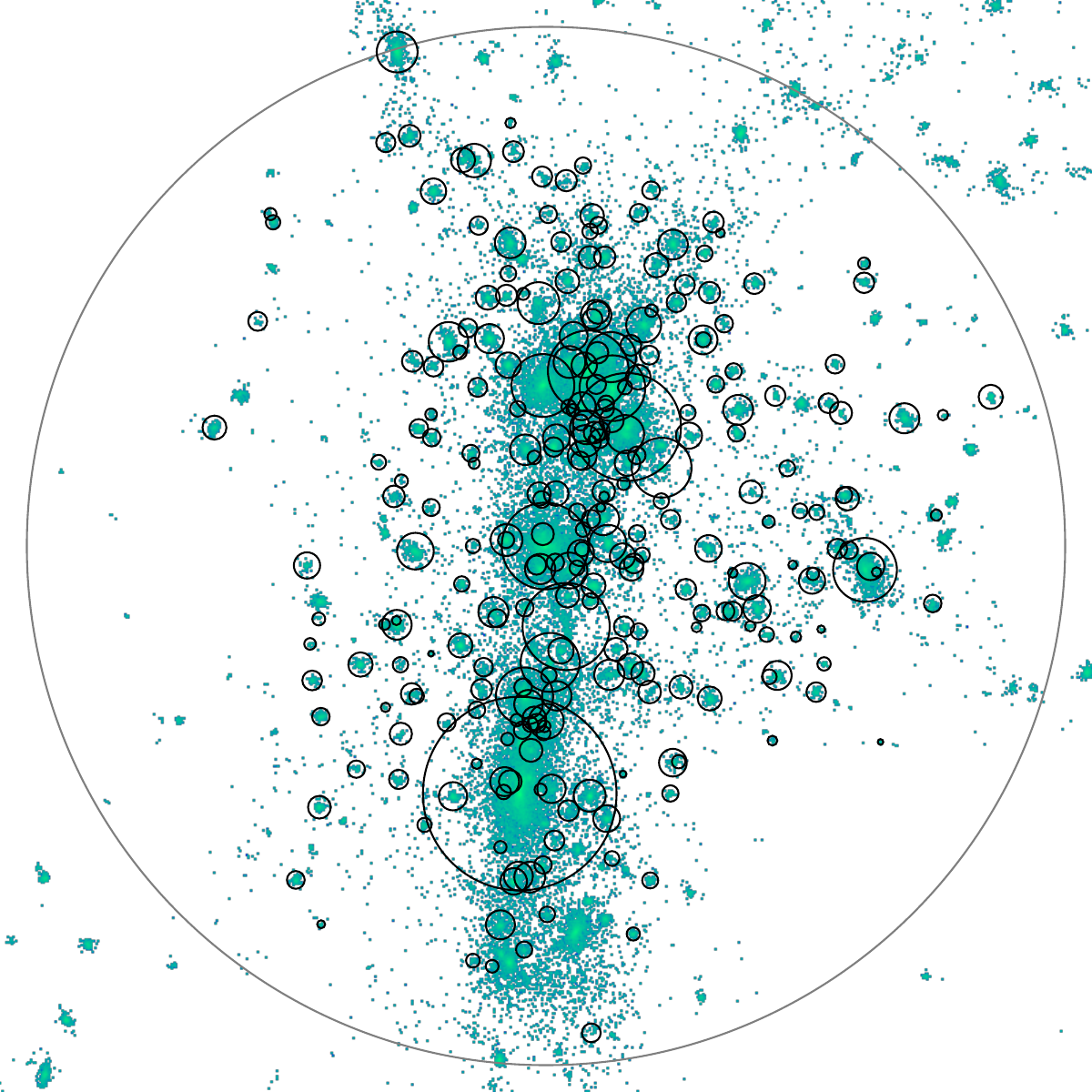}
    \caption{ Visualization of in-situ substructure finding of a large particle cluster from a hydrodynamic simulation using DBSCAN. Image credit: Azton Wells, Argonne National Laboratory.}\label{f:hacc-cluster}
\end{figure}

\Cref{f:hacc-speedup} illustrates the complete measured computational time
(excluding I/O) of the gravity-only simulation over 625 long-range force steps (red curve). As
the simulation progresses, particles cluster, leading to an increase in
solver time. ArborX is utilized in approximately 100 analysis steps during the
simulation, with the most computationally demanding component being the
FOF identification of dark matter halos within the full particle set.

In a series of downscaled test simulations at the same mass resolution, HACC has observed a significant
speedup in performance, ranging from 10 to 12 times faster for FOF finding with ArborX when
compared to a highly optimized CPU OpenMP threaded algorithm. 
\iffalse
Note that this is consistent with the order of magnitude performance improvement reported in the previous section for the challenge problem benchmark. 
\fi
To visually illustrate the impact of such a performance boost in a production run,
\Cref{f:hacc-speedup} features a gray curve conservatively representing a
tenfold increase in FOF execution time. 
This showcases, at a minimum, how slow the solver would have been without ArborX. 
The gray spikes corresponding to the 100 steps are now
distinctly pronounced, clearly highlighting the computational cost of running
analyses without GPU accelerated cluster finding. The speedup ratio in
the bottom panel indicates that even with this conservative estimate, the full
time-stepper shows an improvement of approximately a factor of 2.

The ArborX optimizations are so impactful that HACC can now perform analysis
steps on every long-range force simulation timestep. This capability is particularly important
for hydrodynamic simulations, where cluster finding needs to be executed at a
much higher cadence. Furthermore, ArborX enables the capability of
performing in-situ substructure finding within a simulation. Historically, this
task has necessarily been carried out in post-processing due to its computational expense.

In the case of hydrodynamic simulations, 
gas particles regularly evolve into stars, collectively forming 
galaxies. Each analysis step in HACC requires the identification of
galaxies, achieved by running DBSCAN on all stellar particles using $\minpts = 10$.
\Cref{f:hacc-cluster} visualizes star particles within one of the largest
dark matter halos in the (576 Mpc/h)$^3$ simulation. The largest circle in the
figure encloses the full halo, and each DBSCAN-identified galaxy is marked with
a corresponding circle (using a radius equal to the farthest particle from the
center of the galaxy). Full particle queries of this complexity would have been too expensive to
perform during a simulation prior to the incorporation of ArborX into the
analysis pipeline.

\section{Background}\label{s:background}

\subsection{DBSCAN algorithm}

Here, we give a short overview of the \dbscan algorithm. For more details, we
refer the readers to \cite{ester1996}.

\begin{figure}[t]
  \centering
    \includegraphics[width=0.70\columnwidth]{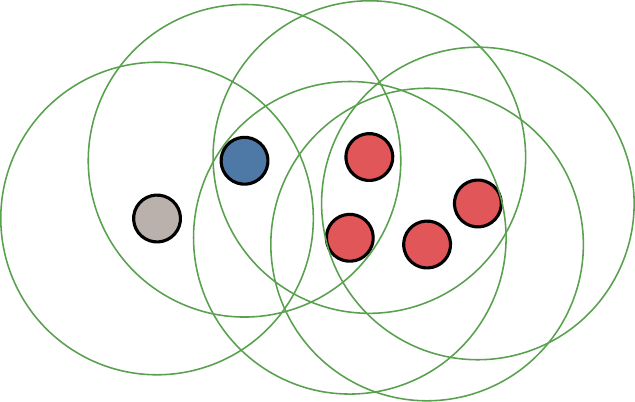}
    \caption{Classification of points for \dbscan with $\minpts = 4$. Core
    points are shown in red, border in blue, and noise are in
    gray.}\label{f:dbscan}
\end{figure}

Let $X$ be a set of $n$ $d$-dimensional points to be clustered. Let $\eps$
and $\minpts$ be two given parameters. An \emph{$\eps$-neighborhood} of a point
$x$ is defined as $N_\eps(x) = \bigl\{y \in X \, | \allowbreak \, dist(x, y)
\le \eps\bigr\}$, where $dist(\cdot,\cdot)$ is a distance metric (e.g.,
Euclidean). A point $x$ with $|N_\eps(x)| \ge \minpts$ is called a \emph{core
point}. A point $y$ is \emph{directly density-reachable}
from a point $x$ if $x$ is a core point and $y \in N_\eps(x)$. A point $y$ is
\emph{density-reachable} from a point $x$ if there is a chain of points $x_1,
\dots, x_n$, $x_1 = x$, $x_n = y$, such that $x_{i+1}$ is directly
density-reachable from $x_i$. Two points $x$ and $y$ are called
\emph{density-connected} if both are density-reachable from some point $z \in
X$. Finally, any point that is not a core point but is density reachable from
one is called \emph{border point}. The remaining points, i.e., the points that
are not core or border points, are called \emph{noise}. Noise points are
considered to be outliers not belonging to any cluster. Any cluster then
consists of a combination of core points (at least one) and border points
(possibly, none). \Cref{f:dbscan} provides an illustration for a set of points
with $\minpts = 4$.

The goal of a \dbscan algorithm implementation is thus to find the individual
clusters in a dataset efficiently.

\subsection{ArborX library}\label{s:arborx}

In this Section, we give a brief introduction to geometric search algorithms
and the ArborX library, setting up a background for performance improvements in
\Cref{s:updates}.

The main index in ArborX is a bounding volume hierarchy (BVH). BVH is a tree
structure created from a set of objects in a multi-dimension space. Each object
is wrapped in a geometric format (\emph{bounding volume}) to form the leaf
nodes of the tree. Each node of a BVH is an aggregate of its children, with the
node's bounding volume enclosing the bounding volumes of its children. The
bounding volume around all objects, called \emph{scene bounding volume},
is stored at the root of the hierarchy.

As was demonstrated in \cite{karras2012}, binary BVH is a good
choice for GPU-based searches, particularly for low-dimensional data
typical in the scientific applications. Fast BVH construction algorithms use a
space-filling curve (Z-curve) to improve the spatial locality of the user data,
followed by a single bottom-up construction to produce a binary tree structure
(hierarchy). While the resulting quality is somewhat worse than produced by the
best available algorithms, the construction procedure is extremely fast and
produces a tree of sufficient quality in most situations.

During the search (also called a traversal), each thread \revision{(a host or a
GPU thread depending on the backend)} is assigned a single query,
\revision{\emph{i.e.}, a spatial or $k$-nearest neighbor search problem}. All
the traversals are performed independently in parallel in a top-down manner. To
reduce the data and thread divergence, the queries are pre-sorted with the goal
to assign neighboring threads the queries that are geometrically close.

To reduce the software development cost, ArborX uses Kokkos \citep{kokkos2022}
for on-node parallelism to allow running on a variety of commodity and HPC
hardware, including Nvidia, AMD, and Intel GPUs. Kokkos abstracts common
parallel execution patterns, such as parallel loops, reductions, and scans
(prefix sums), from the underlying hardware architecture.
In addition, Kokkos provides an abstraction for a multi-dimensional array data
structure called \texttt{View}. It is a polymorphic structure, whose layout
depends on the memory the data resides in (host or device).

Kokkos supports both CPUs and GPUs (Nvidia, AMD, Intel) through providing
backends, e.g., OpenMP, CUDA, SYCL, HIP. Using Kokkos allows running the same
code on CPUs or GPUs by simply changing the backend through a template
parameter, resulting in a higher developer productivity.

Now, we briefly describe the scope of different focus areas within ArborX.

\subsubsection*{Core functionality.}
ArborX supports two kinds of search types: range and nearest. The \emph{range}
search finds all objects that intersect with a query object. For example,
finding all objects within a certain distance is a range search. The
\emph{nearest} search, on the other hand, looks for a certain number of the
closest objects, regardless of their distance. Both searches support
multi-dimensional data, with dimensions ranging 1--10.

These two search types require very different tree traversal algorithms. The
range query has to explore all nodes in a tree that satisfy the given
predicate. It can be implemented in a stackless manner (see
\Cref{s:stackless}). In contrast, the nearest search can terminate early while
it has found the best possible candidates. Nearest search is more complicated
to implement, and relies on a stack and a priority queue structures.

ArborX provides both on-node and distributed (through using MPI)
implementations for both searches.

\ifjournal
\begin{figure*}
  \centering
  \includegraphics[width=0.85\textwidth]{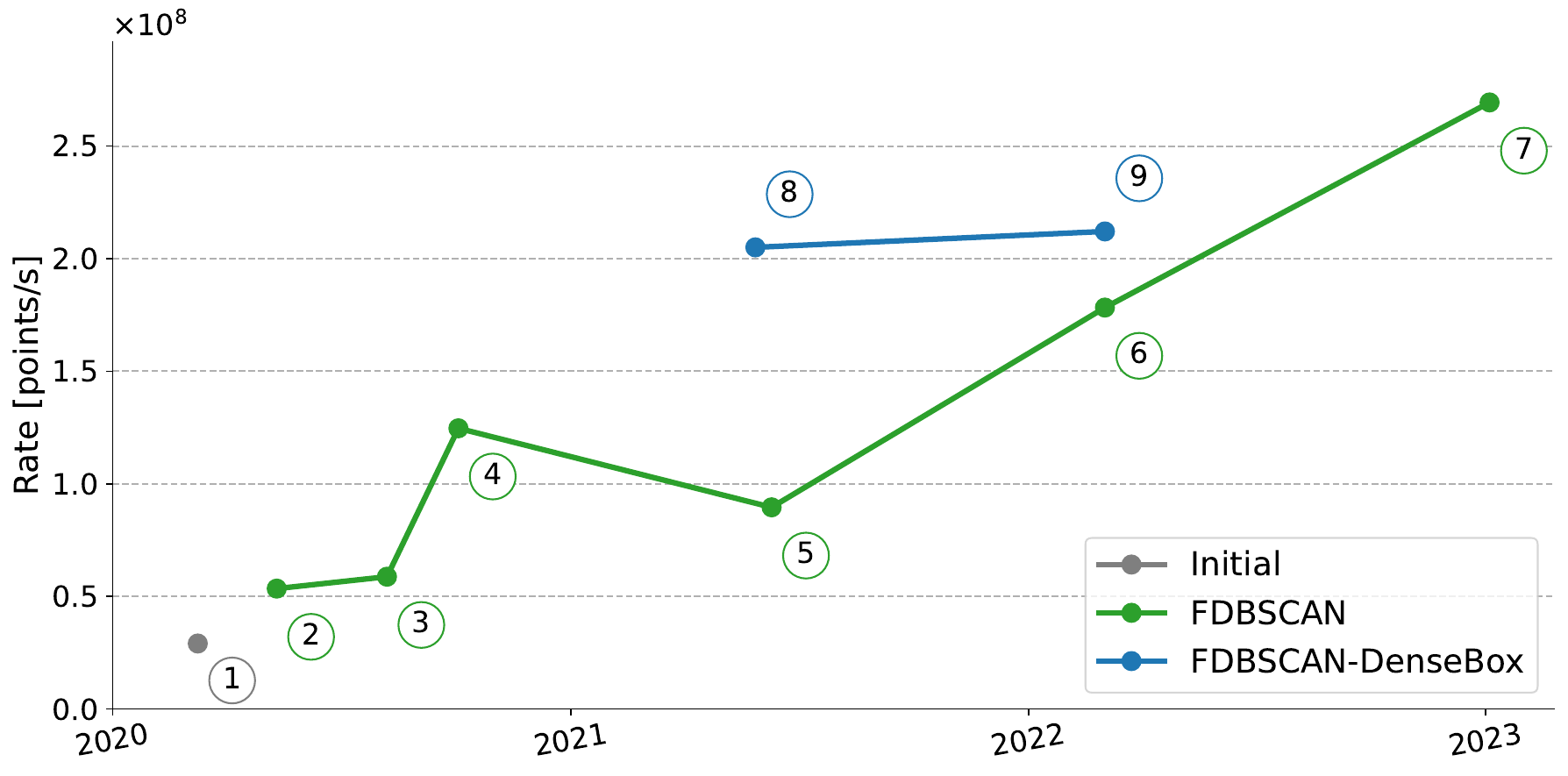}
  \caption{Timeline of the ArborX improvements for the \dbscan benchmark
  problem ($\eps = 0.042$, $\minpts = 2$) on an Nvidia A100 GPU.}\label{f:timeline_eps0.042_minpts2}
\end{figure*}
\fi

\subsubsection*{Clustering functionality.}
ArborX provides implementations for several clustering algorithms that depend
on distance calculations: \dbscan (the focus of this work) and Euclidean minimum
spanning tree \citep{prokopenko2023emst}. There are ongoing efforts to provide
an efficient implementation for the \hdbscan algorithm \citep{campello2015}.

\subsubsection*{Ray tracing functionality.}
ArborX provides basic support for ray tracing.

\subsubsection*{Interpolation functionality.}
ArborX implements moving least squares interpolation algorithm
\citep{quaranta2005}. In this method, support and subsequently the interpolation
operator are constructed through solving local least squares problems defined
by compactly supported radial basis functions.

\section{ArborX improvements}\label{s:updates}
\unless\ifjournal
\begin{figure*}
  \centering
  \includegraphics[width=0.85\textwidth]{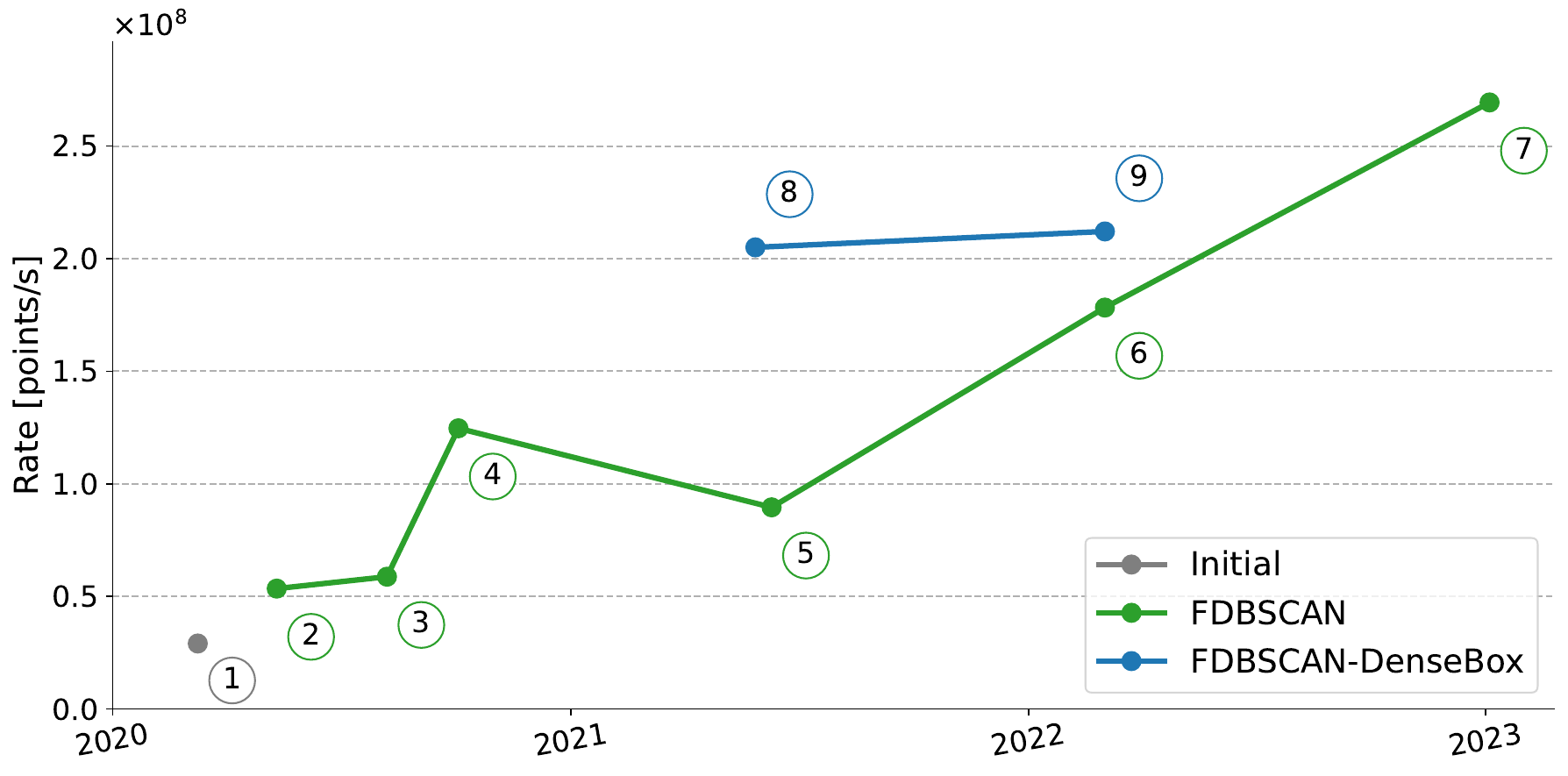}
  \caption{Timeline of the ArborX improvements for the \dbscan benchmark
  problem ($\eps = 0.042$, $\minpts = 2$) on an Nvidia A100 GPU.}\label{f:timeline_eps0.042_minpts2}
\end{figure*}
\fi

ArborX's primary goal is performance. For a user, this comes with an implicit
agreement that ArborX will be diligent in implementing features in a way that
does not slow down user applications. For this to happen, ArborX uses several
benchmarks to test proposed new functionality.

\begin{figure}[t]
    \centering
    \includegraphics[width=0.86\columnwidth]{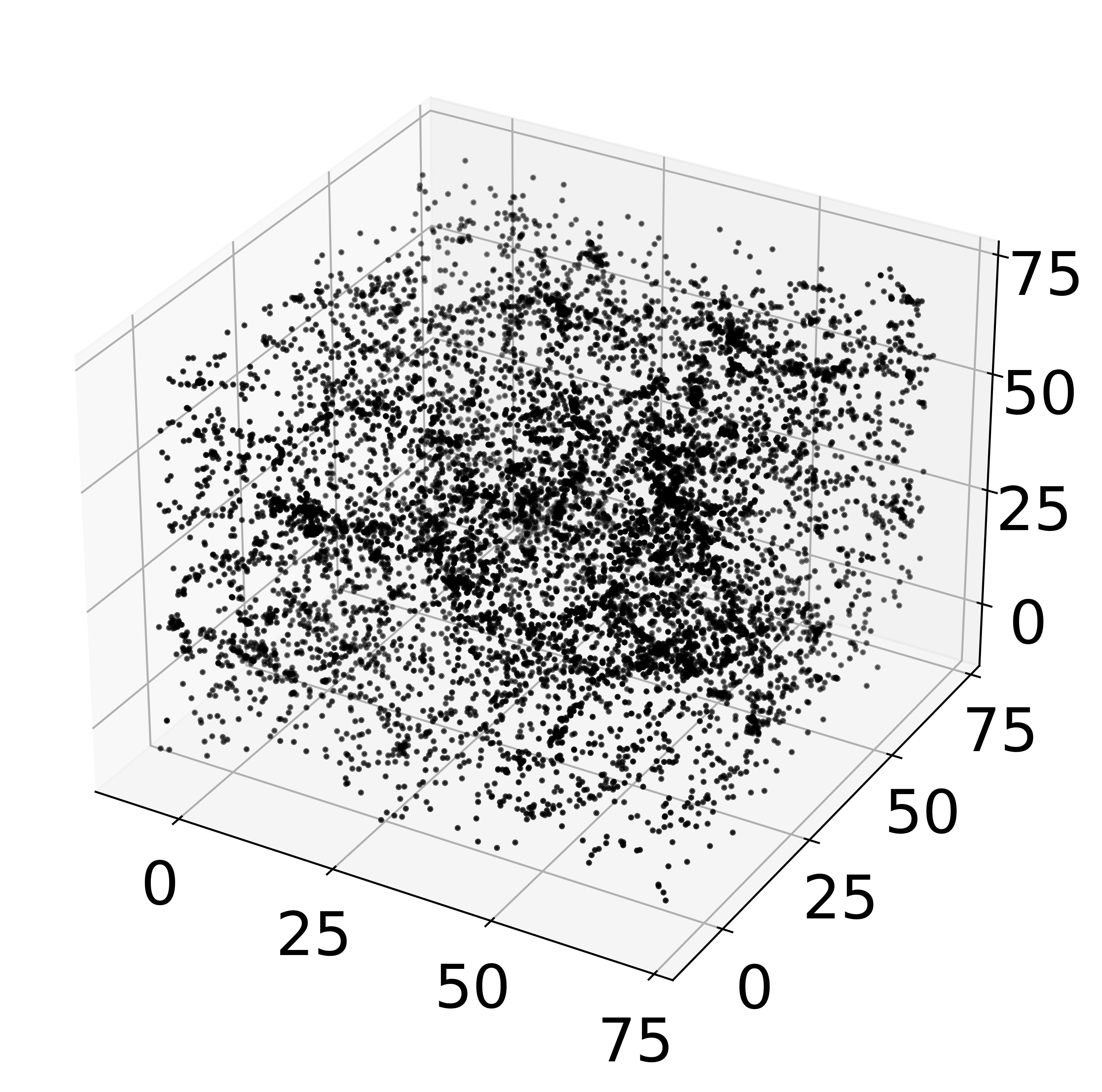}
    \caption{Benchmark problem data sampled from a single rank. The clusters
    are clearly formed.}\label{f:hacc}
\end{figure}

\iffalse
\begin{figure*}
  \centering
  \includegraphics[width=0.82\textwidth]{figures/timeline_saturn_eps_0.5_minpts_2.png}
  \caption{Timeline of the DBSCAN algorithm improvement on the benchmark
  problem on an Nvidia A100 GPU. $\eps = 0.5$, $\minpts = 2$. Green: \fdbscanFDBSCAN. FDBSCAN-Densebox not
  presented, but is close to
  \Cref{f:timeline_eps0.042_minpts2}.}\label{f:timeline_eps0.5_minpts2}
\end{figure*}
\fi

One of these benchmarks is the \dbscan algorithm for a cosmology problem. The
challenge problem for the HACC project is a 12.2 trillion particle
gravity-only simulation with a companion hydrodynamic simulation that models both gas and dark matter with $2\times 1.8$ trillion points. 
A downscaled version of the gas simulation was run using $2\times1024^3$ particles with (256 Mpc/h)$^3$
domain volume. For our dataset, we used a snapshot from the last step of the
simulation performed with HACC, when the clusters are clearly
formed. The data was taken from a single rank of the original 64 MPI rank job, consisting of
$\approx$37M dark matter particles (gas particles are excluded for cluster finding).
\Cref{f:hacc} shows a 3D visualization of the data sample. The value of $\eps$
was set to 0.042\footnote{Following the formula $\eps = b
\left(V/n\right)^{1/3}$, where with $b = 0.168$ is the linking length, $V$ is
the simulation volume ($256^3$), and $n$ being the number particles
($1024^3$)}.

\Cref{f:timeline_eps0.042_minpts2} shows the timeline of the improvements in
ArborX's core search functionality together with the algorithmic developments
for this benchmark problem. The code was run on an Nvidia A100 GPU. Here, we
highlight certain critical points which affected the performance:
\begin{enumerate}
  \item[(1)] Initial implementation (\Cref{s:dbscan_initial})
  \item[(2)] Initial introduction of \fdbscan (\Cref{s:fdbscan}) and use of the callbacks (\Cref{s:callbacks})
  \item[(3)] Switching from using Karras construction algorithm to Apetrei's (\Cref{s:stackless})
  \item[(4)] Switching to using stackless traversal (\Cref{s:stackless})
  \item[(5)] Changing the callback in \fdbscan to improve $\minpts > 2$ case
  \item[(6, 9)] Switching to using 64-bit indices (\Cref{s:64-bit_morton})
  \item[(7)] Using pair-traversal in \fdbscan (\Cref{s:pair_traversal})
  \item[(8)] Initial introduction of \fdbscandense (\Cref{s:fdbscan-dense})
\end{enumerate}
It is clear that the most significant improvements to the performance were the
stackless traversal, the move from 32- to 64-bit Morton codes, and the pair
traversal for \fdbscan. Overall, the performance of the algorithm improved by
factor $\approx$9.2, with the latest version clustering the full $\approx$37M
benchmark problem in under 0.15s on an Nvidia A100. Performance improvements were
mostly monotonic, with the exception for (5), which significantly improved the
\fdbscan callbacks algorithm for $\minpts > 2$ at the cost of a slowdown for
$\minpts = 2$.

Initially, \fdbscandense performed significantly better than \fdbscan, however
the latter became the faster one for this problem with the introduction of the
pair traversal. There is ongoing research to integrate pair traversal into the
\fdbscandense algorithm.

We will now describe each interface, performance and algorithmic improvement in
more details.

\subsection{Interface improvements}

\subsubsection{Callbacks.}\label{s:callbacks}

\begin{figure}
  \centering
  \begin{minted}[fontsize=\footnotesize]{cpp}
struct Callback {
 template<typename Predicate, typename Value>
 KOKKOS_FUNCTION
 RT operator()(Predicate const &predicate,
    Value const &value) const;
};

struct CallbackWithOutput {
 template<typename Predicate, typename Value,
    typename OutputFunctor>
 KOKKOS_FUNCTION
 void operator()(Predicate const &predicate,
    Value const &value,
    OutputFunctor const &output) const;
};
  \end{minted}
  \caption{Callbacks interface. The return type RT could either be \code{void},
  or \code{enum CallbackTreeTraversalControl}. The latter affects the
  traversal, allowing early termination (see
  \Cref{s:early_termination}).}\label{f:callbacks}
\end{figure}

The original ArborX interface was designed to handle the tasks of interest to
the DataTransferKit library. The results were produced as a pair of Kokkos
views (\code{offsets}, \code{values}), with the~\code{values} view containing
the values satisfying the predicates, and the~\code{offsets} view containing
the offsets into~\code{values} associated with each query. However, it was
observed that users are often interested in performing some operation on the
results of each query and not the results themselves. For example, they may be
only interested in the number of neighbors, average distance, or updating some
quantity. In these cases, storing the results may be unnecessary, penalizing
both performance through memory writes and memory usage. For some problems,
storing the found objects results in running out of memory even for simple
counting kernels.

To address this issue, we introduced the callback functionality in ArborX. It
allows execution of a user-provided code on a positive match.
\Cref{f:callbacks} demonstrates the interface. Here, \code{Predicate}
represents a search query (e.g., an object to intersect with), and \code{Value}
represents the data stored in the BVH. We support both \emph{pure} callbacks
that perform a user operation without storing any results, and callbacks that
allow a user to modify the results before storing them.

% In \dbscan, it can be used both in counting the neighbors, as well as merging
% the results using \unionfind.

\subsubsection{Early termination.}\label{s:early_termination}

One of the steps in the \dbscan algorithm is determination of the core points,
i.e., the points that have a specified number of neighbors within the radius.
To find them, we need to execute a traversal procedure.

It is, however, unnecessary to continue the procedure once the threshold has been
achieved and it has been established that a point is a core point. Thus, we
updated the interface to allow a callback to indicate (through a return value)
whether the traversal should continue. For low values of $\minpts$, this saves a
significant amount of time.

There are many other potential applications of this feature. In general, if it
is known that the traversal will produce at most $m$ answers, it can be stopped
after achieving that value. For example, a search for a mesh cell containing a
given particle in particle-in-cell simulations can be terminated after the first
such cell has been identified.

\subsection{Core algorithms improvements}\label{s:core}

In this section, we describe improvements to the interface and several
optimization techniques to improve the performance of the implemented \dbscan
algorithms in \Cref{s:dbscan}.

\subsubsection{Stackless traversal.}\label{s:stackless}

\begin{figure}[t]
  \centering
  \includegraphics[width=0.88\columnwidth]{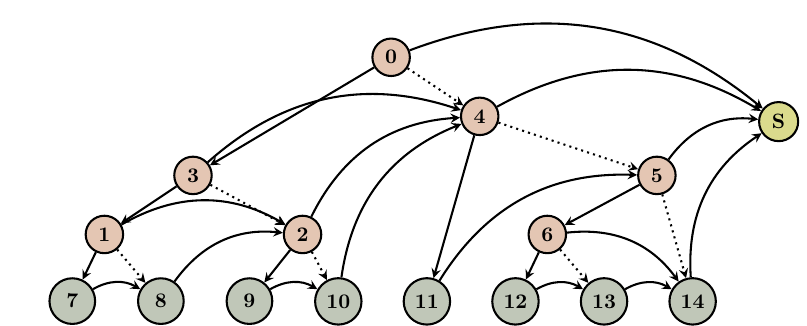}
  \caption{Hierarchy for stackless traversal.
  \revision{
  The references to right children (dotted arrows) of the internal nodes 0-6
  (orange) are replaced with ropes (curved arrows). Additionally, each leaf
  node 7-14, (green) now has a rope. The ropes of the nodes on the right-most
  path point to the sentinel node $S$ (yellow).}}\label{f:karras_ropes}
\end{figure}

Stackless traversal is a technique to avoid explicitly managing a stack of node
pointers for each thread during the traversal. Usage of stacks is undesirable
as it may lead to lower occupancy due to higher memory demands per thread. The
approach in \cite{torres2009} introduced a concept of \emph{rope} (also called
\emph{escape index}), an index of a node where the traversal should proceed
if the intersection test with the current node is not satisfied, or if the node
is a leaf node. In \Cref{f:karras_ropes}, \revision{the references to the right child
(denoted by dotted lines) of the internal nodes are removed, and the ropes
(denoted curved arrows) are introduced for both internal and leaf
nodes}. For
the nodes on the right-most path, the ropes point to the artificial terminal
node called \emph{sentinel}.

Originally, ArborX used Karras' approach \citep{karras2012} for the BVH
construction. The way the Karras algorithm orders internal nodes makes it possible
to set the ropes as part of the standard bottom-up hierarchy construction.

However, to improve the hierarchy construction time, we switched to
using \cite{apetrei2014} as the latter is more efficient. The unfortunate side
effect of the switch was that the ropes had to be set separately, as Apetrei's
algorithm orders internal nodes differently. However, we were able to overcome
this side-effect by finding a way to recover Karras' ordering from the
Apetrei's one.
For more details, see \cite{prokopenko2024stackless}.

\subsubsection{64-bit Morton codes.}\label{s:64-bit_morton}

BVHs are often used in computer graphics for ray tracing, with performance
often compared in terms of rays per second casted for rendering a set of scenes
presented as a set of surfaces (e.g., triangles). One of the challenges of
applying these algorithms to scientific data is the difference in the range of
scales. It is common for the ratio between the densest and the sparsest regions of
scientific data to be several orders of magnitude. This requires certain
adjustments of the algorithms.

Standard \revision{linear BVH}~(LBVH)~\citep{lauterbach2009} algorithms rely on space-filling curves
(typically, Z-order curve
based on Morton codes) to organize the spatial locality of the data. The case
of two objects happen to have the same Morton index resolved in an ad-hoc manner.
However, the scientific data may degrade severely in this case. In our original
implementation, we used 32-bit Morton codes, meaning that each dimension of
the 3D problem was partitioned in 1024 bins. It turned out that due to the very
high density of the particles in some regions, too many particles were assigned
to the same bin, i.e., having the same Morton code. For example, for the
benchmark problem at least 64\% particles
shared their Morton code with at least one other particle, with the maximum of
3,569 having the same index. This resulted in decreased performance of the
algorithms due to the worse hierarchy quality.

We chose to address this problem by using the 64-bit Morton codes. As we can
see in~\Cref{t:64-bit}, showing the statistics for the benchmark problem in, we
eliminate almost all duplicate Morton codes for our problem when using 64-bit
resolution. This becomes even more crucial for larger problems.

Using 64-bit codes has two minor drawbacks. It slightly increases the
hierarchy construction cost due to the slower sorting of the 64-bit integers
compared to the 32-bit ones. Additionally, increasing resolution in Morton
codes will typically result in a deeper hierarchy (in this case, 43 to 49).
However, we found the tradeoff worth it due to significant speedup in traversal
algorithms.

\begin{table}
  \caption{An overview of the hierarchy construction differences between 32-
  and 64-bit Morton codes for the benchmark problem.}\label{t:64-bit}
  \begin{tabular}{lrr}
    \toprule
                                   & 32-bit     & 64-bit    \\
    \midrule
    \#duplicate codes ($>$ 3 times)  & 1,311,912  & 0         \\
    \#points with duplicate code   & 23,539,027 & 528       \\
    max same code duplicates       & 3,569      & 2         \\
    number of hierarchy levels     & 43         & 49        \\
    \bottomrule
  \end{tabular}
\end{table}

\subsubsection{Pair traversal.}\label{s:pair_traversal}

\begin{figure}[t]
  \centering
  \includegraphics[width=0.88\columnwidth]{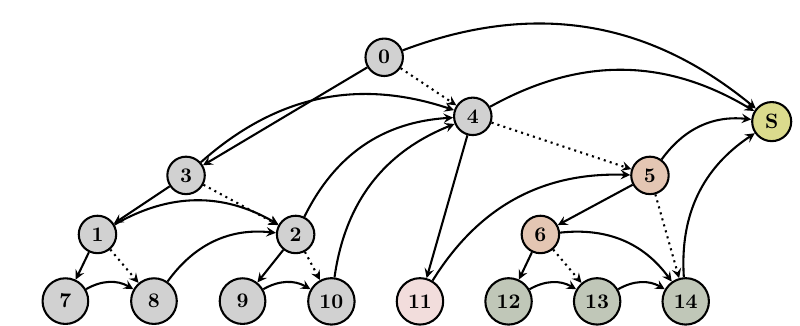}

  \caption{An example of the tree traversal mask for a thread corresponding to
  a point with index 4. \revision{The thread skips all internal and leaf nodes
  in gray, and starts with the corresponding leaf node 14 (pink), which is the
  4-th leaf node. The traversal then follows the standard
  procedure.}}\label{f:pair_traversal}
\end{figure}

Many computational problems, including \dbscan, can be seen as: given a radius
$\eps$, find all pairs $(i, j)$ of points that are within distance $\eps$ of each other,
and execute some operation $F(i, j)$ on the pair. In \dbscan, that operation is
$\union$. In other applications, the operation could be to compute a force
(e.g., in molecular dynamics) or to increase the value of the count (e.g., computing 2-point
correlations).

For this class of problems, each pair needs to be processed only once. Regular
traversal would process each pair twice, once for each thread, resulting in
unnecessary work. Thus, a new approach is desired.

We implemented a new hierarchy traversal algorithm. For a given point $i$,
instead of starting the traversal from the root node 0, we start it with the
leaf node corresponding to this point. Combined with the ropes structure
described earlier, it guarantees that that only pairs $(i, j)$
with $i < j$ will be found, thus processing each pair exactly once.
\Cref{f:pair_traversal} demonstrates the traversal for a thread corresponding
to index 4. It is clear that a thread would need to examine fewer nodes
compared to the regular traversal. This results in fewer memory accesses used
during the traversal, reduced number of distance computations, and reduced
number of applying the operation $F$.

We explained the algorithm above for a tree with ropes. It is possible to
achieve a similar effect for a regular tree with left and right children by
masking subtrees. This would impose certain requirement on the ordering of the
internal nodes, and an initial top-down traversal.

\iffalse
\subsection{Occupancy (50-75\%)}
\fix{What do we want here?}
\fi

\subsection{DBSCAN algorithm improvements}\label{s:dbscan}

\begin{figure}[t]
\captionsetup{labelformat=empty}
\begin{algorithm}[H]
\caption{Disjoint-set \dbscan algorithm}\label{a:dbscan_union_find}
\begin{algorithmic}[1]
\small
\Procedure{DSDbscan}{$X, \minpts, \eps$}
\For {each point $x \in X$}
    \State $N \gets \ttl{GetNeighbors}(x, \eps)$    \label{l:dbscan_union_find:neigh}
    \If {$|N| \ge \minpts$}
        \State mark $x$ as core point
        \For {each $y \in N$}
            \If {$y$ is marked as a core point}     \label{l:dbscan_union_find:core_check}
                \State $\ttl{Union}(x, y)$  \label{l:dbscan_union_find:union1}
            \ElsIf {$y$ is not a member of any cluster}
                \State mark $y$ as a member of a cluster
                \State $\ttl{Union}(x,y)$   \label{l:dbscan_union_find:union2}
            \EndIf
        \EndFor
    \EndIf
\EndFor
\EndProcedure
\end{algorithmic}
\end{algorithm}
\end{figure}

The original DBSCAN algorithm in \cite{ester1996} was hard to parallelize due
to its breadth-first manner of encountering new points. Improvements to the
algorithm in \cite{patwary2012} broke with its breadth-first nature. The
authors used the \unionfind \citep{tarjan1979} approach to maintain a
disjoint-set data structure. The approach relies on two main operations: \union
and \find. $\find(x)$ determines the representative of a set that a point $x$
belongs to, while $\union(x,y)$ combines the sets that $x$ and $y$ belong to.

For completeness, \Cref{a:dbscan_union_find} shows the disjoint-set \dbscan
(\textsc{DSDbscan}) algorithm as proposed in \cite{patwary2012} (Algorithm 2).
Each point only computes its own neighborhood
(\Cref{l:dbscan_union_find:neigh}). If it is a core point, its neighbors are
assigned to the same cluster
(\Cref{l:dbscan_union_find:union1,l:dbscan_union_find:union2}).

We will now describe our initial implementation, and then briefly describe two
newly developed algorithms (for more information, see
\cite{prokopenko2023dbscan}).

\subsubsection{Initial DBSCAN implementation.}\label{s:dbscan_initial}

Our original approach was to solve the special case of \dbscan of $\minpts =
2$. In cosmology literature, it is usually called Friends-of-Friends (FOF). This case
is simpler, as each point either belongs to a cluster as a core point, or is
noise. It is equivalent to finding connected components in the undirected
adjacency graph, with each pair of vertices within $\eps$ of each other have a
corresponding edge in the graph.

Prior to introduction of the callbacks (see~\Cref{s:callbacks}), ArborX only
produced an explicit adjacency graph. Afterwards, we used the
ECL-CC \citep{jaiganesh2018} algorithm to compute the connected components.

The major drawback of this approach was storing the full adjacency graph in
memory. It imposed a severe restriction on the size of the problems one could
run. The memory usage depended not only on the size of the problem $n$, but
also on the parameter $\eps$.

\subsubsection{Reformulated DBSCAN algorithm.}

Both the original \cite{ester1996} and \cite{patwary2012} \dbscan formulations
do not expose enough parallelism for GPU implementations with thousands of
threads. To address that, we reformulated the algorithm to consist of two
phases. In the first phase (\emph{preprocessing}), the algorithm determines the
core points. In the second phase (\emph{main}), it merges the pairs of close
neighbors as they are being discovered.

\begin{figure}[t]
\captionsetup{labelformat=empty}
\begin{algorithm}[H]
\caption{Parallel disjoint-set \dbscan algorithm\label{a:fdbscan}}
\begin{algorithmic}[1]
\small
\Procedure{PDSDbscan}{$X, \minpts, \eps$}
\If {$\minpts > 2$}
  \For {each point $x \in X$ \textbf{in parallel}}
      \State determine whether $x$ is a core point    \label{l:fdbscan:is_core_point}
  \EndFor
\EndIf
\For {each pair of points $x, y$ such that $dist(x,y) \le \eps$ \textbf{in parallel}} \label{l:fdbscan:edges}
  \If {$x$ is a core point}
    \If {$y$ is a core point}
      \State {$\ttl{Union}(x, y)$}
    \ElsIf {$y$ is not yet a member of any cluster}
      \State \textbf{critical section:}
      \State {\hskip1.5em mark $y$ as a member of a cluster}
      \State {\hskip1.5em $\ttl{Union}(x, y)$}
    \EndIf
  \EndIf
\EndFor
\EndProcedure
\end{algorithmic}
\end{algorithm}
\end{figure}

The pseudocode for the reformulated \dbscan (\textsc{PDSDbscan}) algorithm is
shown in \Cref{a:fdbscan}. The preprocessing phase is executed on Lines 3-4.
The \unionfind algorithm is performed on Lines 8 and 11.

The main advantage of the reformulated algorithm is that it executes the
neighbor searches completely in parallel for all data points, allowing it to
take advantage of the underlying parallel index. In addition, it allows
processing each found neighbor as soon as it is found and immediately
discarding afterwards.

\subsubsection{FDBSCAN.}\label{s:fdbscan}

\fdbscan (``fused'' DBSCAN) constructs a search index (BVH) over all the points
of the datasets. Several ArborX improvements described in \Cref{s:core} play a
crucial role in achieving a good performance. We use callbacks to fuse tree
traversal with the counting and \unionfind kernels, avoiding neighbor storage.
This makes the algorithm use $O(n)$ memory. We use early termination in the
counting kernel, stopping a thread once $\minpts$ threshold has been achieved.
In the main kernel constructing the clusters, we then use the pair traversal technique.

\subsubsection{FDBSCAN-DenseBox.}\label{s:fdbscan-dense}

\begin{figure}
  \centering
  \includegraphics[width=0.55\columnwidth]{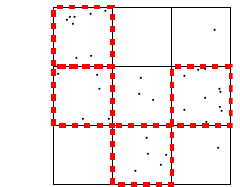}
  \caption{Regular 2D grid with grid size $\eps/\sqrt{2}$ superimposed over
  the dataset. The dense cells for $\minpts = 5$ are shown in red.}\label{f:dense_cells}
\end{figure}

\fdbscandense is a modification to \fdbscan accommodating regions with high
density (with respect to $\eps$). These regions are characterized by the number
of neighbors within $\eps$-neighborhood far exceeding the $\minpts$ value. In
this scenario, many of the distance computations may be avoided.

To achieve this goal, we superimpose a regular grid with the grid cell length
of $\eps/\sqrt{d}$ (with $d$ being the data dimension) on top of the data. The
choice of the length parameter guarantees that each cell's diameter does not
exceed $\eps$. Thus, any grid cell with at least $\minpts$ points in it will
only contain core points, and the distance calculations among them can be
eliminated.. We call these cells \emph{dense}. \Cref{f:dense_cells}
demonstrates an example of such grid over a set of points, with dense cells for
$\minpts = 5$ marked in red.

We modify the BVH construction algorithm of \fdbscan to accommodate dense boxes.
Instead of constructing the hierarchy only on the data points, we construct it
out of a mix of dense cells and points outside of the dense cells. This poses
no challenge to the BVH construction, as it only requires bounding volumes for
a set of objects. During the traversal, if a found object is a dense cell, we
perform a search over all of its contained points.

\section{Conclusions and future work}\label{s:conclusions}

We presented the progress made in ArborX in the past few years to support the
HACC cosmology application. We identified an analysis problem of high
importance to the overall cosmology simulation, and developed and implemented
multiple algorithms to improve the performance of the \dbscan algorithm. We
presented a timeline of the core and algorithmic changes in ArborX using a
benchmark problem. We have also shown that the ArborX speedup, which is of an order 
of magnitude when compared to a highly optimized CPU OpenMP algorithm in HACC, resulted in
a factor of 2 improvement in the full time-stepper performance. 

Our ongoing research has several thrusts. We are developing an efficient
implementation of an advanced density-based algorithm
\hdbscan \citep{campello2015}, which is an improvement over the \dbscan
algorithm. We are working on improving the SYCL based implementation to target
the upcoming Aurora supercomputer. In addition, there are ongoing efforts to
include the auto-tunable interface to select best occupancy parameters during
the simulation runtime.

\ifjournal
\begin{acks}
\else
\section*{Acknowledgements}
\fi
  % ECP disclaimer
  This research was supported by the Exascale Computing Project (17-SC-20-SC), a
  collaborative effort of the U.S. Department of Energy Office of Science and
  the National Nuclear Security Administration.

  % OLCF disclaimer
  This research used resources of the Oak Ridge Leadership Computing Facility at
  the Oak Ridge National Laboratory, which is supported by the Office of Science
  of the U.S. Department of Energy under Contract No. DE-AC05-00OR22725.

  % Argonne disclaimer
  Argonne National Laboratory's work was supported under the U.S. Department of
  Energy contract DE-AC02-06CH11357. Additionally, this study utilized
  resources of the Argonne Leadership Computing Facility, which is a DOE Office
  of Science User Facility supported under Contract DE-AC02-06CH11357. We would
  further like to acknowledge the work of the ExaSky team and the development
  and testing efforts therein.
\ifjournal
\end{acks}
\fi

\ifjournal
  \bibliographystyle{SageH}
\else
  \bibliographystyle{apalike}
\fi
\bibliography{main}

\end{document}